\documentclass[letterpaper,twocolumn,fleqn]{article} 

\usepackage{ist}
\usepackage{amsmath,amssymb,amsfonts}
\usepackage{multirow}
\usepackage{caption}
\usepackage{subcaption}

\title{Turkey Behavior Identification System with a GUI \\ Using Deep Learning and Video Analytics}

\author{Shengtai Ju\textsuperscript{1}, Sneha Mahapatra\textsuperscript{1},
Marisa A. Erasmus\textsuperscript{2}, Amy R. Reibman\textsuperscript{1}, and
Fengqing Zhu\textsuperscript{1}\\
\textsuperscript{1}School of Electrical and Computer Engineering, \textsuperscript{2}Department of Animal Sciences,\\
Purdue University, West Lafayette, Indiana, USA\\}

\date{} 

\begin{document} 

\maketitle 

\thispagestyle{empty} 


\begin{abstract}
In this paper, we propose a video analytics system to identify the behavior of turkeys. Turkey behavior provides evidence to assess turkey welfare, which can be negatively impacted by uncomfortable ambient temperature and various diseases. In particular, healthy and sick turkeys behave differently in terms of the duration and frequency of activities such as eating, drinking, preening, and aggressive interactions. Our system incorporates recent advances in object detection and tracking to automate the process of identifying and analyzing turkey behavior captured by commercial grade cameras. We combine deep-learning and traditional image processing methods to address challenges in this practical agricultural problem. Our system also includes a web-based user interface to create visualization of automated analysis results. Together, we provide an improved tool for turkey researchers to assess turkey welfare without the time-consuming and labor-intensive manual inspection. 

\end{abstract}

\section{Introduction}
\label{sec:intro}
    
Turkey is an important source of poultry meat in the United States and worldwide. Stress and diseases can pose threats to turkey welfare. According to \cite{stress}, stress has direct and negative consequences for turkey welfare. Turkey welfare is critical to turkey production and needs to be monitored closely for optimal production. Videos of turkeys can be used to assess turkey welfare by analyzing their behavior. Currently, researchers and others studying animal behavior need to spend a lot of time watching video recordings of turkeys in order to identify their behavior, which is both time-consuming and labor-intensive. There is an urgent need to develop an automated system to assist turkey researchers to conduct turkey welfare research. Such a system should be able to track turkeys accurately and detect changes in their behavior, and provide data visualization and analysis for user-friendly interaction. 


To determine whether a turkey is sick or not, we need to identify different types of behavior. Behaviors of interest include: walking, stationary/sitting still, eating, drinking, preening, beak pecking, and feather pecking. Healthy and sick turkeys will engage in these behaviors with different frequencies and duration. For behaviors such as walking and being stationary, we can estimate each individual turkey's distance traveled within a fixed time frame, which requires the knowledge of the turkey's location in each video frame from a turkey tracker. 
For behaviors like eating and drinking, we need to locate each turkey and determine whether it is eating or drinking based on its proximity to the feeder or drinker. When computing the proximity to the feeder or drinker, both the turkey's location and its corresponding head location are required to make an accurate estimation of its behavior. Therefore, in addition to the turkey tracker, a turkey head tracker is also needed. 

Existing studies that focus on monitoring animal behavior and health often require the use of wearable devices or are solely based on the researcher's observation, which can be intrusive to the turkeys and time-consuming to the humans. For example, Stevenson \textit{et al.} \cite{stevenson2019validity} use accelerometers to monitor changes in turkey behavior. Results show that habituation to wearing accelerometers on turkey legs greatly impact the validity and reliability of data. In another study, turkey behavioral responses are recorded by human observers \cite{erasmus2014temperamental}. 

\begin{figure}[htb!]
\centerline{\includegraphics[width=\linewidth]{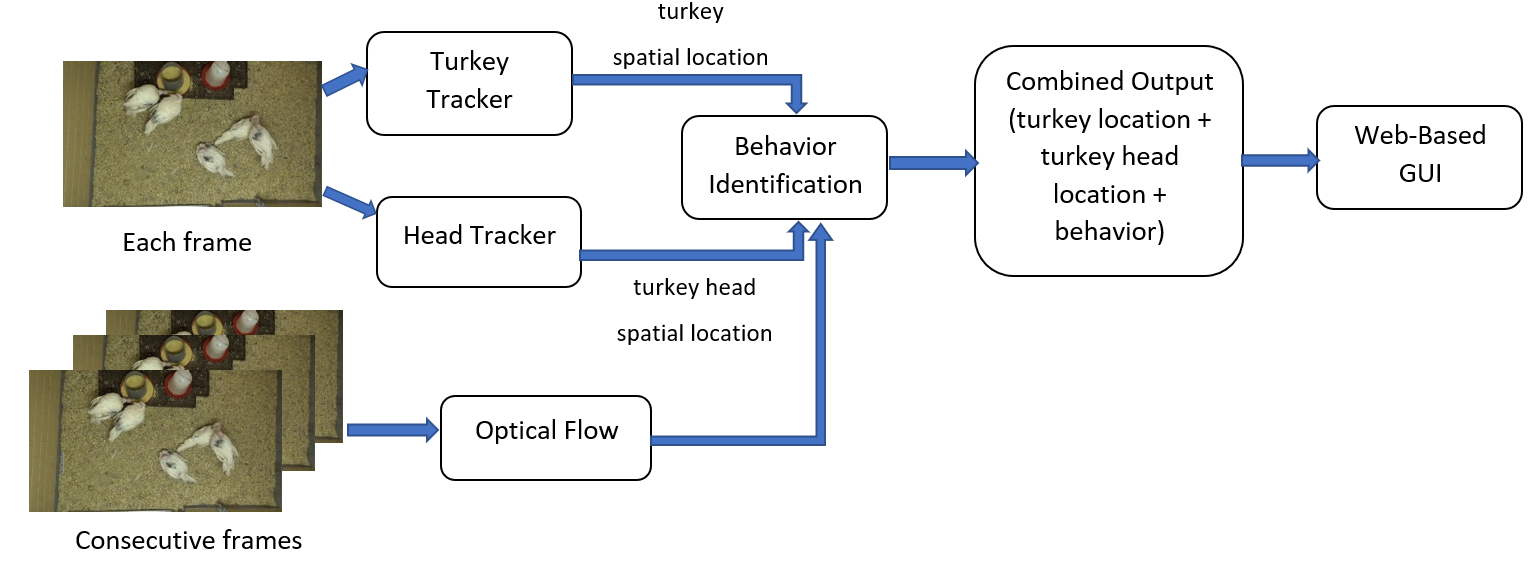}}
\centering
\caption{Overview of proposed system.}
\label{block_diag}
\end{figure}

To better analyze turkey behavior and reduce labor, we propose an automated turkey tracking system with four major components: a turkey tracker, a turkey head tracker, a turkey behavior identification module, and a graphical user interface (GUI), as shown in Figure \ref{block_diag}. The turkey tracker is used to track the location of each turkey in every frame, indicated by a bounding box around each turkey during the tracking process. The turkey head tracker is used to determine the location of each turkey's head in every frame and is also indicated by a bounding box around each turkey's head. By combining information about turkey location and turkey head location, the behavior identification module detects different behaviors of interest and the duration and frequency of each behavior. Our system also includes a GUI to assist the turkey researchers with visualization of the video analysis results.  Our GUI can generate different interactive plots that enable users to focus on individual turkey behavior during custom time intervals as well as aggregated behavior for the entire video. Our main contributions include: 1) applying deep learning and video analytics to create a practical turkey behavior identification system, 2) creating an interactive web-based GUI to visualize automated video analysis results of turkey behavior, and 3) developing a turkey head tracker based on color histograms.

\section{Related Work}
\label{sec: related_work}

In our earlier work \cite{modified_csrt}, we implemented a turkey tracker based on the CSRDCF \cite{csrdcf} tracker, which uses correlation filters to track objects. Recently, deep-learning based object trackers have shown promising results for many practical problems. In particular, DeepSort \cite{deepsort} uses a deep-learning based object detector combined with the Kalman filter \cite{kalman} to carry out multi-object tracking, which is the task of tracking a set of objects in a sequence of frames \cite{fiaz2018tracking}. The object detector generates detection results as bounding boxes, which are used to initialize the trackers. The Kalman filter predicts the next state of each tracked object. The tracker then associates newly detected bounding boxes with predicted locations from the Kalman filter using the Mahalanobis distance metric, which is effective for computing the distance between a point and a distribution. Another component of the DeepSort tracker is an appearance feature extractor. The appearance feature extractor computes the appearance features for each detected bounding box and associates the detection to one of the existing tracks by computing the cosine distance between two distributions. The overall distance metric is a weighted sum of the Mahalanobis and cosine distances. In our application, we choose the DeepSort tracker because it is efficient and more suitable for video with many frames. For object detection, YOLOv3 \cite{yolov3} is a widely used object detector and has shown promising performance in many applications. Other object detectors such as the R-CNN \cite{rcnn}, and the Fast R-CNN \cite{frcnn} generate different regions and assign class labels to each region. They can be slow because each region needs to be classified. YOLO looks at the entire image and detects objects within the image using a single neural network. Therefore, it is faster and more efficient.  

Several object tracking systems have been designed to track animals. ToxTrac \cite{toxtrac} and BioTracker \cite{biotracker} are two recent animal trackers with GUI. ToxTrac \cite{toxtrac} is an open-source object tracking software for animals such as fish and insects. It includes several different algorithms for tracking, such as thresholding, background subtraction, and Kalman filtering. The GUI in ToxTrac can show individual tracks as well as some statistical information. BioTracker \cite{biotracker} is similar to ToxTrac in terms of software functionalities. It includes tracking algorithms such as the background subtraction tracker, and Lukas-Kanade optical flow tracker. The GUI is used to load video files and run the tracker. It can also be used to visualize tracking results and statistics of tracks. These existing systems do not provide the components that are needed for our application including the tracking of turkeys and their heads, identifying individual turkey behavior and built-in interactivity of the GUI. 



\section{Methods}
\label{sec: methods}
\subsection{System Overview}
\begin{figure}[htb!]
\centerline{\includegraphics[width=0.5\linewidth]{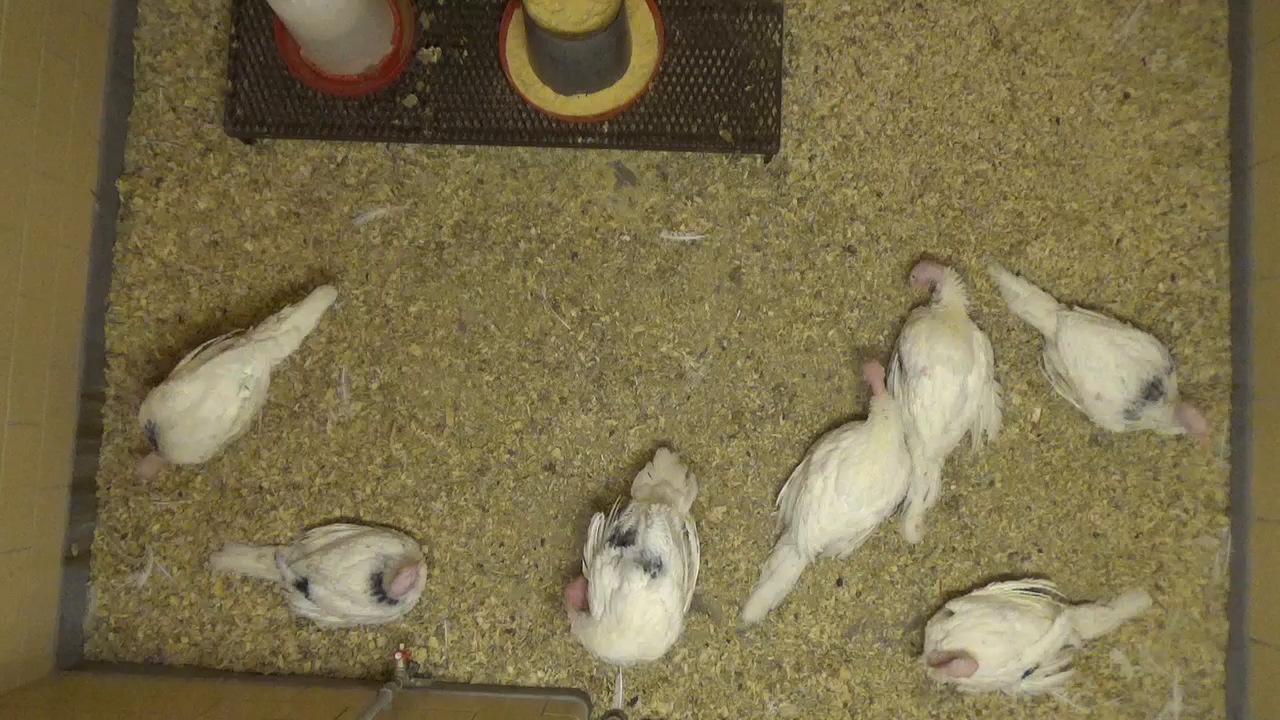}}
\centering
\vspace{.05in}
\caption{Experimental room with seven turkeys}
\label{room_7t}
\end{figure}

Our method consists of a turkey tracker, a turkey head tracker, a turkey behavior identification module, and a GUI as described in Figure \ref{block_diag}. The input to our system is a video clip. For each frame, the DeepSort tracker predicts the location of each turkey by drawing a bounding box around it. The tracker also assigns a unique ID to each turkey so that an individual track can be established. The head tracker uses color information in different color spaces to find the heads of all tracked turkeys. Dense optical flow is also applied to two consecutive frames to detect motion at a fine level. Based on information from the aforementioned components, the behavior identification module identifies the type of behavior associated with a specific turkey. Results from the trackers and behavior identification are imported into the GUI along with the video clips to generate different statistical plots. 
Figure \ref{room_7t} shows an example frame of the turkey video. As can be seen from the figure, turkeys look highly similar and it can be challenging to identify the heads because of background color. In the following sections, we describe each component of our system: 1) turkey tracker, 2) head tracker, 3) behavior identification module, and 4) GUI.

\subsection{Video Analytics}
\subsubsection{Turkey Tracker}
The DeepSort tracker \cite{deepsort} is used to predict the location of each turkey in every frame. The tracker relies on the YOLOv3 detector \cite{yolov3}, which is fully trained on our custom turkey dataset. Details about the training data is presented in the dataset and evaluation metrics section. 
A turkey track contains the spatial location and bounding box information of each turkey in a frame. It can be used to generate the temporal path of turkeys in the video, whereas a turkey tracker is used to describe the overall tracking algorithm. For a turkey track to be initialized and established, detection of that turkey has to be successful for a fixed number of frames. After a track is established, the tracker uses a Kalman filter to predict the next state of the track. The predicted state of an existing track from the Kalman filtering is matched with the new detection results by computing the Mahalanobis distance. This association accounts for short-term motion of the object \cite{deepsort}. Appearance information of an object also needs to be considered because after long-term occlusion, a turkey should be re-identified by the tracker with the same ID that was originally assigned to that turkey. Distance between the newly detected turkeys and previous turkey instances is computed by cosine distance in the feature space. A CNN model pre-trained on a person re-identification dataset \cite{mars} is used to extract features. Even though this model was trained on data for re-identifying humans, it still performs well for extracting turkey features. By incorporating both motion and appearance information, the DeepSort tracker provides accurate predictions of where the turkeys are in each frame. Compared to the original implementation as described in \cite{deepsort}, we modified a few parameters to achieve the best results for our application. We changed the parameters inside the Kalman filter so that it only tracks the position variables without considering the aspect ratios of the bounding boxes. We also increased the number of successive detections needed to create a new track. These changes were made because turkeys move differently compared to humans, and shape deformations such as wing flapping and fast walking can cause false positive detection.

\subsubsection{Turkey Head Tracker}
To accurately find the heads of the turkeys, we focus on the color information of the heads. As shown in Figure \ref{room_7t}, it can be challenging to distinguish the turkeys' heads from the background due to color similarity and small size. To make the heads more distinguishable from the background, we apply contrast enhancement to the frames as a pre-processing step. By converting each frame from RGB to HSV color space and by weighing the \textit{S} channel more, we increase the contrast of every frame processed by our head tracker. 
Since the color of turkey heads vary in different videos due to lighting conditions, we use manual initialization of the turkey heads location to achieve best results. We manually initialize turkey heads by drawing bounding boxes around them at the beginning of the video once the turkey tracks have been established. In subsequent frames, we look for patches around all turkey heads that best match the one initialized in the first frame. We also incorporate target updating, lost target detection, and lost target recovery throughout the entire video.

The manual initialization of each head is a $25\times25$ pixels patch. For each target patch, we compute its color histograms in two different color spaces: HSV, and CIELAB. Histograms are computed for each individual color channel, for a total of six histograms from both color spaces. During the initialization stage, the location of each head target is stored. In subsequent frames, the head tracker looks for the head in a tight search region constrained by the initialization. In order to account for motion of the head, we expand the search region of the head to be $50\times50$ pixels. By dividing this window into overlapping $25\times25$ regions, we compute the six color histograms for each patch and compute the cosine distance between each patch and the original target. A weighted distance measure is used to compute the final distance metric, as shown in Equations \ref{disthsv} and \ref{distlab}. The sum of these two distances is used as our final distance measure. The overlap of different patches is set to be 5 pixels. 
\begin{equation}
\mathrm{distHSV} = \begin{bmatrix} \mathrm{distH} & \mathrm{distS} & \mathrm{distV} \end{bmatrix} \begin{bmatrix} 4/5 & 1/10 & 1/10 \end{bmatrix}^T
\label{disthsv}
\end{equation}
\begin{equation}
\mathrm{distLab} = \begin{bmatrix} \mathrm{distL} & \mathrm{dista} & \mathrm{distb} \end{bmatrix} \begin{bmatrix} 2/5 & 2/5 & 1/5 \end{bmatrix}^T
\label{distlab}
\end{equation}

Since turkeys are free to move around the room, there will be slight changes to the color of turkey heads because of different lighting in different areas of the room. Therefore, we need to update the turkey-head targets regularly to make our head tracking more accurate. Turkey-head targets are updated every 3 seconds in our experiments. Since error can accumulate throughout the tracking process and target updating, there is a need to detect when a turkey head is lost. Here, error comes from the identified best matches containing many background pixels and drifting away from the actual head center. The cosine distance between the current head patch being tracked and the target patch is used to determine if a head is lost. Prior knowledge on turkey body shape is also used when determining whether a head is lost or not, since we know that a turkey's head cannot be too far away from its body. The threshold for identifying a lost head is determined ad hoc based on the characteristics of the video. 

\subsubsection{Behavior Identification system}
This section presents the behavior identification module, which relies on the estimated location of turkeys and their heads. We focus on detecting the following three behaviors: walking, eating, and drinking. Walking can be detected by calculating the distance traveled by each turkey within a fixed amount of time. If the distance traveled by a turkey exceeds a fixed threshold, we can label this turkey as walking. For detecting eating and drinking, we need to mark the locations of the feeder and drinker first. By using the turkey locations and turkey head locations, we can measure each turkey's proximity to the feeder and drinker. We then make a decision based on whether the turkey is closer to the feeder or drinker. Optical flow is used to detect more complicated behavior such as preening and aggressive interactions. The output of the behavior identification module is the duration and frequency of each behavior. The minimum separation between two instances of the same behavior is three seconds. For example, if a turkey pauses walking for less than three seconds and continues to walk, it is still considered to be walking during the pause. 

\subsection{Web-Based GUI}
We designed a web-based GUI to allow easy and efficient visualization and interaction with video analytics output. We used four different formats to present the data: graphical interpretation, video interpretation, animation and analysis, and tabular data. The web-based GUI interface is written in Python, and uses the Dash and Plotly \cite{dash_plotly} libraries. 

\subsubsection{Graphical Interpretation}
The graphical interpretation tab allows the user to view the data with different types of graphs in two main categories: comparison graphs and statistical graphs. The main purpose of the comparison graphs is to allow the user to compare results of different turkeys. For example, the activity level of turkeys can be estimated based on distance traveled by each turkey in the distance vs. time plot. The spatial location graph can be used to visualize where inside the room each turkey spends most of its time, to help understand its behavior. We include graphs to show both individual and aggregate analysis since changes in behavior can be reflected both individually and as a group. Statistical graphs are designed to provide basic statistical analysis of the turkeys' movement. Examples include histogram distance (PDF) plot, cumulative distribution function (CDF) plot, violin plot \cite{violin_plot}, and 2D kernel density estimation plot.
There are also many interactive features allowing the user to connect the graphical data with the associated video data. Features include selecting which turkeys to view, comparing user-selected turkeys, clicking on a data point to view the location in the video where the data point occurred, zooming into certain sections of the graph, and selecting certain sections of the graph.

\subsubsection{Video Interpretation}
In the Video Interpretation tab, the user can look at the video of the turkeys being tracked in much more detail such as changing speed, updating intervals to focus on, and seeking to certain parts of the video (first quarter, half, third quarter, and end). 

\subsubsection{Animation and Analysis}
\begin{figure}[htb!]
\centerline{\includegraphics[width=0.8\linewidth]{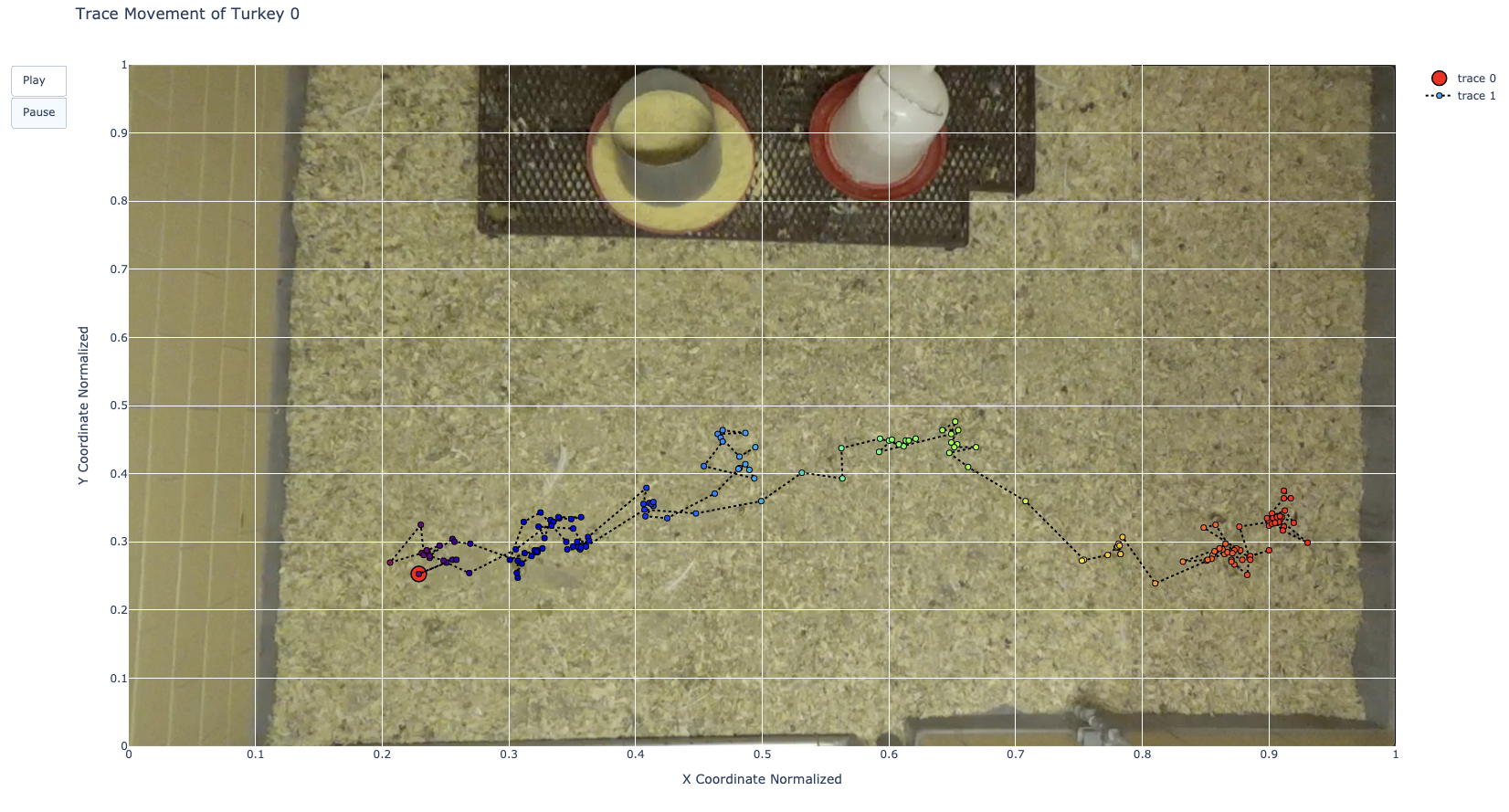}}
\centering
\caption{An example of the animation and analysis plot for one turkey. A play and stop button is provided to visualize the turkey movement as each point represents the turkey position in a frame.}
\label{AniAndAny}
\end{figure}

In the Animation and Analysis tab, the user can view the movements of a single turkey as a point object. This allows the user to see how the turkeys move every second. Figure \ref{AniAndAny} shows an example of the animation and analysis plot.

\subsubsection{Tabular Data}
In the Tabular Data tab, the user can view the data provided by the trackers. The Tabular data provides a 5-column list where each row represents a turkey every second, and the five columns represent the turkey's ID, x-coordinate of the centroid, y-coordinate of the centroid, width of the bounding box, and height of the bounding box. Users can highlight certain boxes, or sort the data numerically in either ascending or descending order.  

\section{Experiments and Results}
\label{sec:experiments}
\subsection{Dataset and evaluation metrics}
\label{ssec:data}
Our dataset contains videos of multiple turkeys in small pens. All turkeys are white commercial turkeys raised to similar weeks of age so that they have similar color and sizes. Videos are captured using commercial grade cameras at HD resolution ($1280\times720$) and 30 frames per second (FPS). Turkeys are kept in experimental pens at the Purdue College of Veterinary Medicine. Inside each pen, there is one feeding station and drinking station. All turkeys are free to move around the room. To evaluate our system, we picked two 3-minute video clips for evaluation. The first clip has five turkeys inside the room whereas the second clip has seven turkeys inside the room. Both clips are taken from a top-down camera viewing angle but in different rooms with different lighting conditions. These 3-minute video clips demonstrate system performance over a long period of time. With longer evaluation clips, we can observe the impact of lighting condition changes on turkey bodies and heads as they move around the room. We also separate these two 3-minute clips into six 1-minute clips to test our system performance over shorter periods of time. With shorter clips, we can observe the turkey behaviors more closely and identify which portion of the longer clips are more challenging for tracking turkeys. 

We trained two YOLOv3 models for turkey detection using our custom turkey images. The first model is trained on 432 images with five turkeys in each image. The second model is trained with all 432 images used in training the first model, plus an additional 486 images with seven turkeys in each image for a total of 918 training images. The first model is applied to videos with five turkeys in the pen while the second model is applied to videos with seven turkeys in the pen. Testing data is never included in the detector training data. 

Since our system tracks multiple objects, we adopt the commonly used multi-object tracking (MOT) evaluation metrics to evaluate our turkey tracker and turkey head tracker. The general and widely-used MOT metrics contain many terms, most of which are not relevant to our application. For our system, we chose to use the multiple object tracking accuracy (MOTA) and multiple object tracking precision (MOTP) proposed in the CLEAR MOT metrics \cite{clearmot} as shown in Equations \ref{mota} and \ref{motp}, respectively. $m_t$, $fp_t$, and $mme_t$ are the number of misses, the number of false positives, and the number of mismatches at time $t$, respectively. $g_t$ is the number of objects present at time $t$. $d_{t}^{i}$ is the distance between a matched object in the ground truth with its corresponding hypothesis, at time $t$. $c_t$ is the number of matches, at time $t$. Since MOTA subtracts three error ratios from the constant one, it can actually take on negative values. MOTP measures the distance error for matched objects and their hypotheses. MOTA is reported as percentage and MOTP is reported as pixels per match. Note that MOTA increases with improved performance while MOTP decreases. 

\begin{equation}
    MOTA = 1 - \frac{\sum_{t}(m_{t} + fp_{t} + mme_{t})}{\sum_{t}g_{t}}
\label{mota}
\end{equation}
\vspace{-.1in}
\begin{equation}
    MOTP = \frac{\sum_{i,t} d_{t}^{i}}{\sum_{t}c_{t}}
\label{motp}
\end{equation}

For evaluating our behavior identification module, we adopt commonly used activity recognition metrics mentioned in \cite{ar_eval}. Specifically, our behavior detection is evaluated based on events, not frames. We evaluate the eating, drinking, and walking behavior in terms of precision, recall, number of insertions, and number of deletions.  Precision is the ratio between true positives and total number of detection. Recall is the ratio between true positives and total number of objects present. An insertion (\textit{I}) is defined as a detected event with no corresponding event in the ground truth. A deletion (\textit{D}) is defined as failure to detect an event. 
We also compute the intersection over union (IOU) ratio between the detected behavior time intervals and the ground truth time intervals. IOU is computed as a ratio, with the intersection between two time intervals in the numerator and the union of two time intervals in the denominator. 


\subsection{Results and Discussion}
\label{sec: results}
\subsubsection{Turkey tracker}
As can be seen from Table \ref{table:tracker_results}, our DeepSort tailored to tracking turkeys performs better than our previous turkey tracker \cite{modified_csrt}, under all evaluation clips for both short-term and long-term tracking. The clip ID column contains information about the clip length and the number of turkeys inside the experimental room. For example, \textit{3m5t} represents a 3-minute video with five turkeys inside the room. Our previous turkey tracker uses the CSRDCF tracker\cite{csrdcf} and YOLOv3\cite{yolov3} for detection. The increase in MOTA is dramatic for our newly proposed tracker. The pixel error for matched turkeys is less than 20 pixels for all evaluation clips. 

\subsubsection{Turkey head tracker}
For the turkey head tracker, we also compare our method against our previous tracker \cite{modified_csrt}. Table \ref{table:tracker_results} shows that our head tracker outperforms our previous tracker in all evaluation clips. MOTA values are greatly increased while MOTP values are reduced. Some MOTA values for \cite{modified_csrt} are negative because the number of misses and false positives are greater than the total number of ground truth objects. Negative MOTA values indicate that the tracker fails to track meaningful objects for a large fraction of the evaluation clip. 

\begin{table}[]
\small
\begin{tabular}{|c|c|c|c|c|c|}
\hline
\multirow{2}{*}{Clip ID} & \multirow{2}{*}{Method} & \multicolumn{2}{c|}{Turkey Tracker} & \multicolumn{2}{c|}{Head Tracker} \\ \cline{3-6} 
                                                                                        &                         & MOTA $\uparrow$             & MOTP $\downarrow$           & MOTA $\uparrow$             & MOTP $\downarrow$          \\ \hline
\multirow{2}{*}{$3m5t$}                                                              & \cite{modified_csrt}                  & 47.6\%            & 21.22           & -51.0\%          & 24.92          \\ \cline{2-6} 
                                                                                        & Ours                    & \textbf{98.7\%}   & \textbf{10.65}  & \textbf{48.7\%}  & \textbf{17.76} \\ \hline
\multirow{2}{*}{$3m7t$}                                                              & \cite{modified_csrt}                  & 10.7\%            & 21.62           & -50.4\%          & 29.47          \\ \cline{2-6} 
                                                                                        & Ours                    & \textbf{91.6\%}   & \textbf{13.88}  & \textbf{63.2\%}  & \textbf{17.78} \\ \hline
\multirow{2}{*}{$1m5t^{1}$}                                                              & \cite{modified_csrt}                  & 57.3\%            & 23.45           & 14.9\%           & 22.46          \\ \cline{2-6} 
                                                                                        & Ours                    & \textbf{99.0\%}   & \textbf{10.72}  & \textbf{76.3\%}  & \textbf{14.82} \\ \hline
\multirow{2}{*}{$1m5t^{2}$}                                                              & \cite{modified_csrt}                  & 54.8\%            & 23.56           & -3.8\%           & 19.49          \\ \cline{2-6} 
                                                                                        & Ours                    & \textbf{99.0\%}   & \textbf{10.72}  & \textbf{53.9\%}  & \textbf{17.06} \\ \hline
\multirow{2}{*}{$1m5t^{3}$}                                                              & \cite{modified_csrt}                  & 39.4\%            & 23.34           & -47.6\%          & 29.87          \\ \cline{2-6} 
                                                                                        & Ours                    & \textbf{98.5\%}   & \textbf{11.84}  & \textbf{26.2\%}  & \textbf{13.31} \\ \hline
\multirow{2}{*}{$1m7t^{1}$}                                                              & \cite{modified_csrt}                  & 12.3\%            & 20.47           & -15.3\%          & 27.26          \\ \cline{2-6} 
                                                                                        & Ours                    & \textbf{91.3\%}   & \textbf{12.52}  & \textbf{70.7\%}  & \textbf{19.36} \\ \hline
\multirow{2}{*}{$1m7t^{2}$}                                                              & \cite{modified_csrt}                  & 40.3\%            & 21.44           & 26.6\%           & 23.89          \\ \cline{2-6} 
                                                                                        & Ours                    & \textbf{93.9\%}   & \textbf{16.00}  & \textbf{50.2\%}  & \textbf{16.06} \\ \hline
\multirow{2}{*}{$1m7t^{3}$}                                                              & \cite{modified_csrt}                  & 5.6\%             & 28.50           & 16.7\%           & 21.24          \\ \cline{2-6} 
                                                                                        & Ours                    & \textbf{89.6\%}   & \textbf{13.92}  & \textbf{66.6\%}  & \textbf{17.61} \\ \hline
\end{tabular}
\caption{Turkey tracking and turkey heads tracking results (bold numbers are best results for each clip and each task.)}
\label{table:tracker_results}
\end{table}

\subsubsection{Behavior identification module}
Given an ethogram that defines different turkey behavior, we labeled ground truth behavior data for walking, eating, and drinking in terms of time intervals for one of our evaluation clips with 7 turkeys. The ground truth contains data for 4 out of 7 turkeys inside the room. However, our system fails to detect one of the turkeys consistently and only outputs the location information for a small fraction of the frames. We omit this turkey for behavior evaluation since we cannot reliably estimate its locations for a consecutive sequence of frames. Our goal is to successfully detect as many actions as possible and not miss any action. Therefore, recall is more important than precision for our application. Among the 3 turkeys, the ground truth contains a total of 10 occurrences of walking, 13 occurrences of eating, and 1 occurrence of drinking. As shown in Table \ref{table: behavior_id}, our behavior identification module achieves over 60\% in recall and over 53\% IOU in time intervals for correct detection. 
\begin{table}[]
\small
\tabcolsep=0.08cm
\begin{tabular}{|c||c|c|c|c|c|}
\hline
         & Precision & Recall & \# of \textit{I} & \# of \textit{D} & IOU \\ \hline
Walking  &   60\%    &  60\%  &      4               &      2               & 0.6015 \\ \hline
Eating   &  73\%  & 85\%& 3                    &       1              & 0.5319 \\ \hline
Drinking &  20\%     &  100\% &      4               &    0                 & 0.6923 \\ \hline
\end{tabular}
\caption{Turkey behavior identification results}
\label{table: behavior_id}
\end{table}

\subsubsection{GUI}
\begin{figure}[htb!]
\centerline{\includegraphics[width=0.8\linewidth]{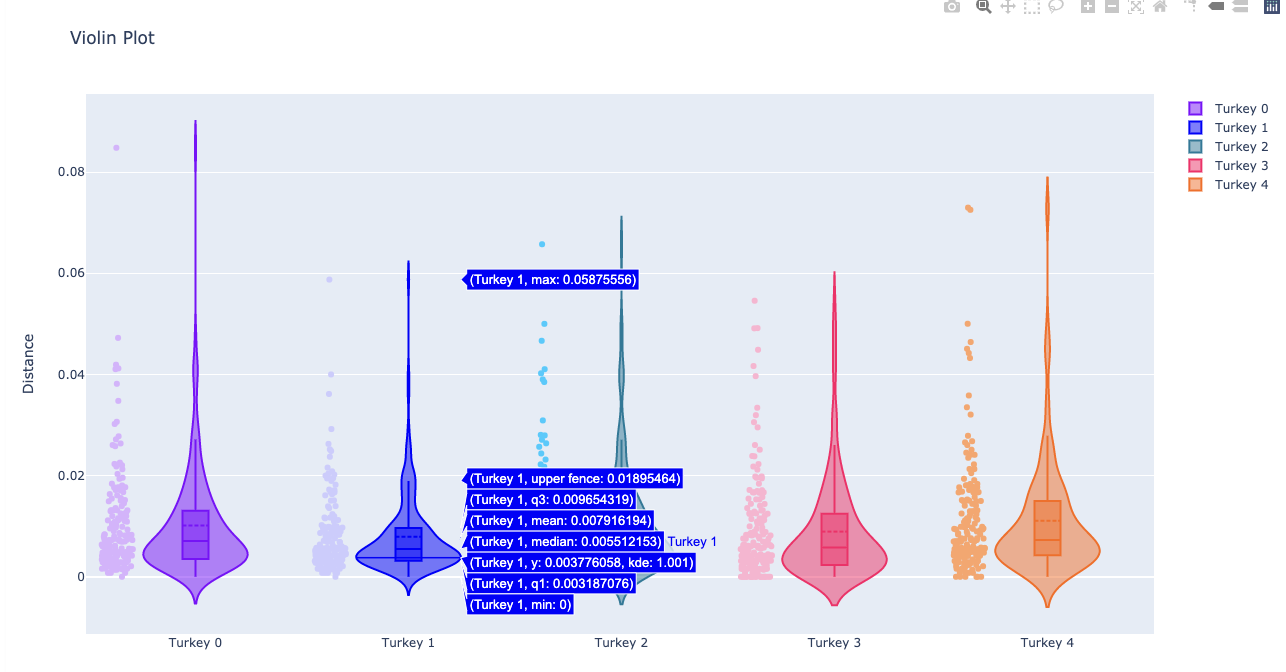}}
\centering
\caption{Violin Plot for Turkey 1 with characteristics for Turkey}
\label{violinPlot}
\end{figure}

\begin{figure}[htb!]
\centerline{\includegraphics[width=0.6\linewidth]{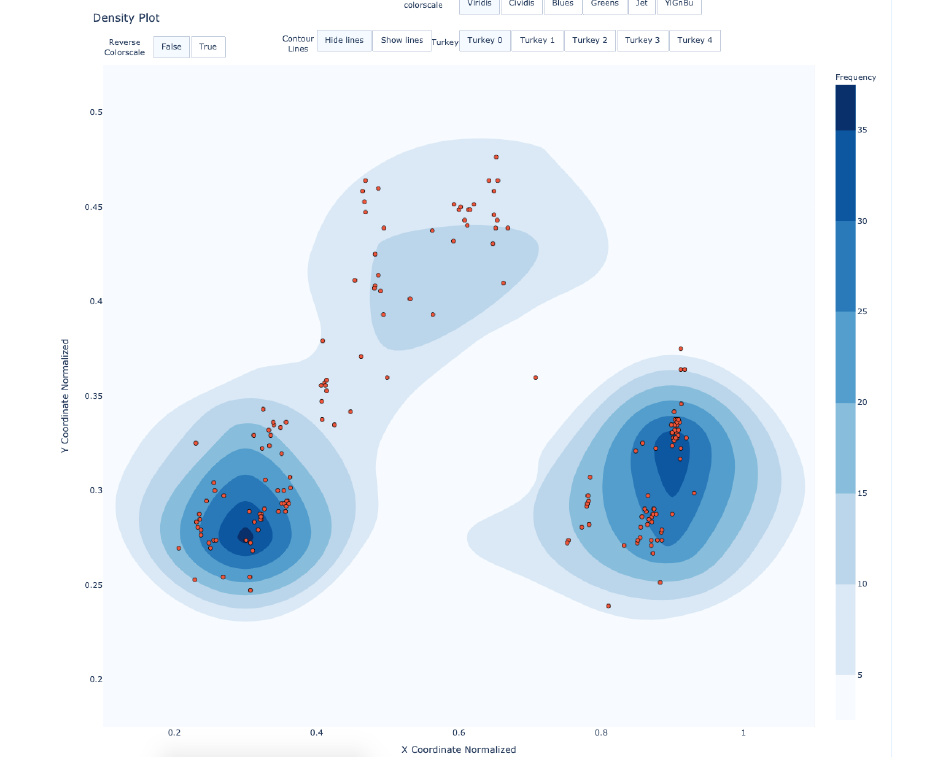}}
\centering
\caption{2D Density Plot of Turkey Locations.}
\label{2d_density}
\end{figure}

\begin{figure}[htb!]
\centerline{\includegraphics[width=0.8\linewidth]{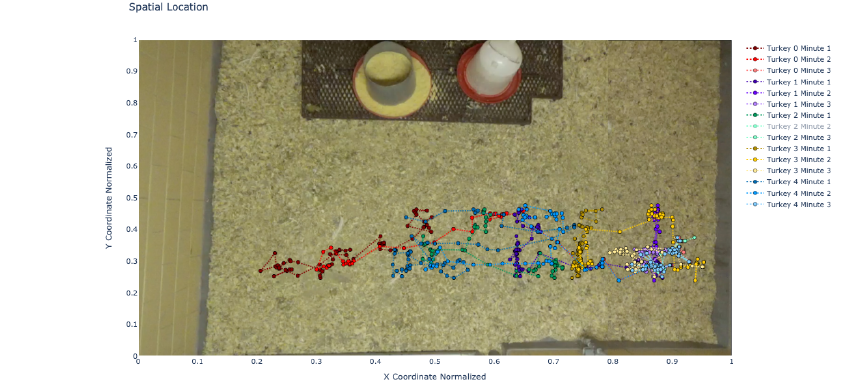}}
\centering
\caption{Turkey Trajectory Plots Color Coded by Time.}
\label{turkey_track}
\end{figure}

\begin{figure}[htb!]
\centerline{\includegraphics[width=0.8\linewidth]{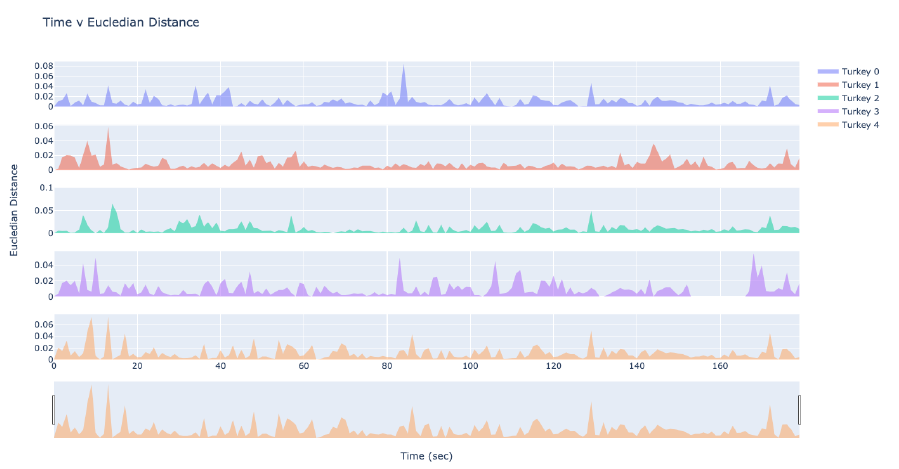}}
\centering
\caption{Individual Turkey Spatial Distance with an Interactive Time Window.}
\label{ind_turkey}
\end{figure}

A few example plots are shown here to demonstrate the functions of the GUI. 
Figures \ref{turkey_track}, \ref{ind_turkey}, and \ref{2d_density} are examples of the comparison plots. Figure \ref{violinPlot} is an example of the statistical plots. Figure \ref{turkey_track} is a plot of the turkey trajectory, color coded by time intervals and turkey ID. Figure \ref{ind_turkey} shows that the user can use our GUI to select a custom time window and focus on an individual turkey to visualize results. Figure \ref{2d_density} is an example of turkey locations plotted as a 2D density plot. The darker color means that the turkey is spending more time at that location inside the room. Figure \ref{violinPlot} shows the violin plot for distance traveled by each turkey. The violin plot is similar to a box plot but instead it also represents the rotated kernel density plot (showcasing concentrated places of data points). These plots enable researchers to engage in both individual and aggregate behavior analysis and provide researchers with abundant interactivity with the tracking results.

\section{Conclusion and Future Work}
\label{sec: conclusion}
In this paper, we introduce a video analysis system that tracks turkeys and turkey heads, identifies turkey behavior, and provides an interactive GUI to help visualize the results. We demonstrate good tracking performance on several evaluation video clips both for our DeepSort-based \cite{deepsort} turkey tracker and our color histogram based turkey head tracker. Our turkey tracker and head tracker perform significantly better than our previous method \cite{modified_csrt}. Results from our behavior identification module also show that we can accurately detect each behavior and capture the duration of the activity. Our interactive GUI provides different forms of data visualization to interpret the tracking results. For future work, we will evaluate our system on more video data to test its robustness under different conditions. Also, according to \cite{erasmus2018welfare}, injurious pecking, such as head pecking and feather pecking, are major behavioral issues for commercial turkeys. We plan to expand the behavior identification module to incorporate a more comprehensive collection of behaviors, especially the aggressive interactions between turkeys, so that abnormal turkey behavior can be identified and treatment can be applied in a timely manner. 





\small

\bibliographystyle{IEEEtran}
\bibliography{ref}

\begin{biography}
Shengtai Ju is a second-year PhD student of Electrical and Computer Engineering at Purdue University, West Lafayette, Indiana.  Shengtai received his B.S.E.E in Electrical and Computer Engineering from Purdue University in 2019, with highest distinction.

Sneha Mahapatra is a recent graduate of Purdue University with a BS degree in Computer Engineering. Her interests lie in Computer Vision and Software Engineering. Attending grad school in the Fall, she will major in Computer Science where she will focus her efforts in Machine Learning and Computer Vision.

Marisa Erasmus is an assistant professor in the Department of Animal Sciences at Purdue University. Work in her laboratory focuses on developing animal-based measures of welfare for chickens, turkeys and ducks, among others, and examining the effects of environmental and management factors on animal welfare. She received her BS and MS degrees from the University of Guelph in Canada and her PhD from Michigan State University. 

Amy R. Reibman has been a Professor at Purdue University in the School of Electrical and Computer Engineering since 2015, where she focuses on video analytics for real-world applications like animal agriculture and food safety. Prior to that, she was an Inventive Scientist at AT\&T Labs doing research in video transmission and video quality assessment. She is a Fellow of the IEEE.

Fengqing Zhu is an Assistant Professor of Electrical and Computer Engineering at Purdue University, West Lafayette, Indiana. Dr. Zhu received the B.S.E.E. (with highest distinction), M.S. and Ph.D. degrees in Electrical and Computer Engineering from Purdue University in 2004, 2006 and 2011, respectively. Her research interests include image processing and analysis, video compression and computer vision. Prior to joining Purdue in 2015, she was a Staff Researcher at Futurewei Technologies (USA). 
\end{biography}

\end{document}